# Effectiveness of Intrusion Prevention Systems (IPS) in Fast Networks

Muhammad Imran Shafi, Muhammad Akram, Sikandar Hayat, and Imran Sohail

**Abstract**—Computer systems are facing biggest threat in the form of malicious data which causing denial of service, information theft, financial and credibility loss etc. No defense technique has been proved successful in handling these threats. Intrusion Detection and Prevention Systems (IDPSs) being best of available solutions. These techniques are getting more and more attention. Although Intrusion Prevention Systems (IPSs) show a good level of success in detecting and preventing intrusion attempts to networks, they show a visible deficiency in their performance when they are employed on fast networks. In this paper we have presented a design including quantitative and qualitative methods to identify improvement areas in IPSs. Focus group is used for qualitative analysis and experiment is used for quantitative analysis. This paper also describes how to reduce the responding time for IPS when an intrusion occurs on network, and how can IPS be made to perform its tasks successfully without effecting network speed negatively.

——————————— ◆ ———————————

## 1 INTRODUCTION

EASIER connectivity to different networks of computers is a major reason of recent advancements in networks and communication infrastructures. More and more companies are making huge investments into online applications and services. People are also keen to make use of these online systems for getting more convenience and ease of use. In addition of easier and cheaper availability of Internet, research and heavy financial investments have made networks very fast. Fast networks and highly available Internet facility is a best combination for online business activities. More bandwidth is available now for online applications making them faster and offering higher data rates. Heavy volume of network can contain any kind of malicious contents that can destroy integrity and credibility of our information systems. According to [3], for any business organization; information is as important as capital and loss or theft of information can bring unbearable financial consequences to the organization.

Although fast networks have given rise to business activities on Internet, they have also introduced a whole new range of computer related threats never known before. These threats are related to security issues and cyber crimes. In recent years there are a huge number of online applications under attack from viruses, mallware, malicious contents, Trojans and many more. Types of attacks include network packets sniffing, network invasion, IP spoofing, email spoofing, password attacks and disclosure of secret information. Yahoo, Google, AltaVista, Overture and MSN are all big IT organizations that remained under attack in recent years. All these events show the seriousness of the situation. Although Internet is open and free for all financial organizations, banks and customers, at the same time it is also open for script kiddies and black hat coders. Hackers have organized themselves into virtual communities. They have their established ways of information and experiences sharing. They have all resources and expertise to attack on any online resource. This alarming situation demands a high emphasis on security requirements for all online resources.

Many organizations have developed different techniques to meat the threat but no technique comes fully up to expectations. Firewalls, IDPSs, proxy servers and honey pots are the examples of anti-hacking techniques. In recent years, emphasis has been shifted to IPSs for being better of the lot. In addition to preventing the network attack, an IPS can create logs of all events, send alerts, initiate additional steps, send automated emails and make phone calls. They can also be made to identify compound attacks and can learn about unknown future attacks. These qualities give them a prominent position in defense strategies.

## 2 BACKGROUND AND RELATED WORK

Malicious traffic is increasing in computer networks both in terms of volume and diversity of attacks. Newer and more complex attacks are being used by hackers to exploit network resources. Hackers are no more dump and dependent upon some automated tools. Script kiddies are capable of producing smart codes to make systems behave their way. This situation motivates the need of a system that can enhance the security level of networks significantly. There can be many possible solutions for the problem but IPSs are the best of breed due to many rea-

————————————————

- *Muhammad Imran Shafi is SEO of SamSoft software-house and visiting Lecturer in Punjab University, Pakistan.*
- *Muhammad Akram is Lecturer in College of Computer Science & Information Systems, Najran University, Saudi Arabia.*
- *Sikandar Hayat is Student Administrator in Blekinge Institute of Technology, Sweden.*
- *Imran Sohail is Assistant professor in Fatima Jinnah Woman University, Pakistan.*

sons. They are quite capable of detecting and preventing diverse types of intrusion attempts and can be further improved in their performance and accuracy. Without detection and prevention systems, network resources are wide open for intruders. Traditional security measures can not cope with the situation elegantly and full intrusion prevention cannot be achieved.

Intrusion prevention is a complete process of identifying and preventing malicious contents within the network traffic going into or out of a network [4]. From above we understand that a network based IPS sits at the gateway to network, intercepts the network traffic, detects the malicious contents within traffic and takes immediate steps to stop the attack efforts. Intrusion attempts are done by hackers to exploit the resources of organizations and making huge benefits.

Organizations are investing huge sums of financial and other resources to protect their network systems and information resources but at the same time, number of intrusion attempts is increasing with the speed more than ever before. A successful intrusion attempt by any hacker to any organization not only compromises the secrecy of information and causes financial loss but most important damage done is loss of credibility and trust. People will no longer like to do business with an organization that cannot protect its information resources.

Different technologies are being used by IT organizations including anti-virus programs, firewalls, proxy-servers, IPS etc to ensure the protection of their valuable information resources. Although no technology provides the answers of all security related questions but IPSs being better than other contending technologies have gained professionals trust in recent years. IPSs take proactive approach to defend network resources. IPSs use attack signatures to identify known attacks and identify malicious behavior of data contents being entered the network. IPSs also perform network protocols analysis to protect against protocol violations [10].

Although IPSs give rise to security level of a network significantly but in case of networks providing high bandwidth and huge data rates, their performance diminishes. Most IPS today target to flow aggregation and reconstruction of TCP streams but they are really unable to handle gigabit rate links [5]. Increased data rate demands more system resources to analyze it including memory, bandwidth and computability. Availability of such resources to IPS from hosts decreases their capability for handling other services. IPS needs special design for handling high data rates. Intruders are also aware of this situation. They are well equipped with resources to circumvent traditional IPS. They enhance the traffic from different systems to victim system to keep it busy in handling huge data sets and then launch more sophisticated attacks that are harder to detect and defend. Intruders often benefit from spoofing to hide their identities. This makes it difficult to stop the attack. When a host identifies an intrusion attempt from some system it blocks it IP address but what actual have to do only is to spoof to some other address and continue attacking the victim.

Distributed IPS capable of handling gigabit links is the solution for said problem. IPS components should be specialized hardware implemented and distributed across the network to cover all the entry points to network.

## 3 EXPERIMENT AS QUANTITATIVE APPROACH

"*An action that has various outcomes that occur unpredictably and can be repeated indefinitely under same conditions*" [12]. An experiment is a good technique for scientific study and research. Experiment can be performed to produce data, results, to check validity of underlying theory or to check correction of available data. Also this technique is flexible to use with systems and human beings. In field of network security, experiments are done with configuration and installation point of security software to enhance their effectiveness. Different products with varying configurations are tested against different attack types and effort is made to find a proper security product with a configuration that gives maximum rise to security level of network.

Experiments have many advantages of other testing techniques including better control of researcher over test variables, generation of numeric data (many statistical tests possible) and possibility of experiment replication. Experiment is the best scientific way to find relationship between cause and effect. An experiment includes definition, planning, operation, analysis and interpretation, presentation and package [13].

### 3.1 Definition
Experiment is done to compare the effectiveness of IPS in is software based centralized and hardware based distributed implementations. Another query that is going to be answered is the configurations of IPSs that offer maximum resistance to intrusion attempts on network. High speed network traffic along with malicious contents will be used as input for these IPS implementations to check their effectiveness.

On the basis of data produced during this experiment, research questions of this paper will be answered.
1. Objects: Objects of this experiment are network traffic (with different malicious contents in it), network throughput, software-based and hardware-based implementations of IPSs.
2. Purpose: Purpose of experiment is to compare the effectiveness of IPS in software based implementation and hardware based implementation. Other purpose is to find the configurations of these implementations that offer maximum resistance against intrusion attempts.
3. Quality Focus: Quality focus of experiment is finding the configuration of IPS at maximum traffic throughput in a fast (gigabit) network.
4. Perspective: Experiment is done according to network managers and network security experts' perspective. This experiment is beneficial for network managers and all kinds of computer based security experts. Intrusion prevention companies can also benefit this experiment to enhance their products performances for high speed, high bandwidth networks.

5. Context: Experiment is done using two open source IPSs one with software based centralized implementation and other with hardware based distributed implementation with slight modifications as no current implementation is designed for high speed networks.

## 3.2 Planning

1. Context: Context of this experiment is comparison of two IPS implementations to find the better choice with best possible performance. Experiment has following characteristics:
   - Experiment is conducted on data flowing inward and outward of a network gateway.
   - Network and software intrusions is the problem faced by all computer world including application developers, operating system vendors, Internet Service Providers (ISP), security products developers, governments, banks, academic and financial institutions. This experiment has potential to effect many areas software development and maintenance.
   - Nature of experiment is online as it is conducted on live data.
2. Hypothesis and Formulation: Null Hypothesis (H0): Increased network throughput and variations in intrusion attempts do affect the IPS's efficiency negatively.
   
   H0: Prevention of intrusion attempts (normal throughput and average intrusion attempts in a unit time) = Prevention of intrusion attempts (high speed, high throughput and enhanced intrusion attempts with increasing complexity)
3. Variable Selection: Variables for this experiment are carefully selected including network throughput, intrusion attempts and degree of prevention for intrusion attempts.
   
   Network throughput is independent variable. Intrusion attempts and their complexity is also an independent variable.
   
   Degree of prevention for intrusion attempts is a dependent variable whose value depends upon the success of IPS in its job. This experiment is done to find the IPS type and its configuration that gives maximum degree of prevention for intrusion attempts.
4. Subjects: IPSs used in this experiment are the subjects. These IPSs are chosen due to their open source nature. Implementations of IPSs are modified to work in distributed environment with high bandwidth networks. Data collection agents in IPS are distributed in nature whereas data analysis and prevention is done by a centralized manager.

## 3.3 Design

1. Randomization: In order to ensure randomization, data packets being traveled on the network are chosen randomly using an open source packet generator. Packet generator is capable of generating data as it is generated in live communication over network. It is also ensured that all kinds of intrusion threats are also passed to network along with clean data to check how successful IPS is in its task of intrusion prevention. All data passed to IPSs is logged.
   
   Second factor (network throughput) is also kept random through same packet generator. Throughput is kept random but its value is always kept near the upper bound of network throughput capacity.
2. Blocking: Network throughput is always kept in the range of maximum throughput capacity of network so that performances of IPSs can be checked under maximum pressure and load.
3. Balancing: Experiment is balanced by passing same data to both IPS implementations. Data passed to software-based centralized IPS is extracted from log files and same data is passed to hardware-based distributed IPS.
4. Design Type: Design type of experiment is two factors with two treatments. Scenario of an academic institution network is considered that is directly related to topic under discussion. This high speed network stores student profiles, academic data, research work, books etc and is accessible to outer world through Internet, telnet and ssh. Compromising any one source on network can further compromise other resources causing financial, academic, information and trust loss among different communities.
5. Instrumentation: Instruments used during this experiment include IPS components, random data generator, data-logging and extracting mechanism and events storage mechanism. Since data generation, data logging/extraction and events (intrusion attempts, prevention actions) storage is done automatically, no human training is required. Only factor chosen with human interaction is value of network throughput.
6. Validity Evaluation: Validity of experiment is done to check how successfully it fulfills its aimed objectives. Validity is also done to ensure its scientific significance. According to [13], validity of experiment is done under heads given below as:

Conclusion Validity: "*Conclusion validity is the degree to which conclusions we reach about relationships in our data are reasonable*" [14]. A threat to this experiment can be the definitions of intrusion attempts. IPS intrusion database is populated with only known threat types. It is possible that one intruder invents a method never known before. In that case it can never be assured that such attacked can be identified or not.

Internal Validity: Internal validity refers to treatment of independent variables and the outcome generated by experiment.

In this experiment, network throughput and data nature should not prevent IPSs from doing their job. One threat to this validity can be the amount of time IPS takes to identify an intrusion attempts as some intrusion attempts are too time specific and before their activation, it is not possible to identify them (Chernobyl Virus).

External Validity: External validity refers to the quality

of generalization for experimental outcomes.

One threat to this study can be the number of false positives. Keeping IPS rules strict (always recommended), IPS suffers this problem. Other threat can be network architectures. Many older networks may not support IPS installation.

Construct Validity: "*Construct validity refers to the degree to which inferences can legitimately be made from the operationalizations in your study to the theoretical constructs on which those operationalizations were based*" [14].

High performance of IPS under maximum throughput and with variations in intrusion attempts refers to the construct validity of this experiment as same outcome was expected from this experiment.

### 3.4 Operation

Operation refers to actual execution of experiment.

Preparation: Preparations are done before actual execution of experiment. Both software-based and hardware-based IPSs are installed on network gateway one by one to perform their job. It is ensured that no data passing to network bypasses IPS data collection point. A 400 GB hard disk is reserved to store network logs and events and a same size hard disk is installed for backup purposes. Backup of log data is taken after every 24 hours. An automated email/SMS/phone call system is also installed to generate automated alerts. A special utility capable of shutting down all network services is also installed that activates only when network is compromised through intrusion attempt.

Execution: Experiment is conducted with both IPSs independently on same network scenario of educational institution. Automated data generator is configured to continuously generate and send data on network for 72 hours. Same data is used as input for both IPS operations to check their relative performance. Different intrusion types are used along with clean data in a fast network environment to ensure relevance of experiment results. Attack types used during experiment include Denial-of-Service attack (DoS), Nimda Virus, Code Red, SirCam, ExploreZip, Chernobyl Virus, Trojans, Opas, BugBear, Blaster, Polip, SQL Slammer etc.

Data Validation: Data used in this experiment is taken from real network scenarios. Simple attacks and attack variations are used along with clean data to test IPSs in best possible way.

Experiment is repeated many times with varied inputs. Different data entry points are used for data without bypassing data collection points for IPSs.

### 3.5 Analysis and Interpretation

Descriptive Statistics: Throughput of data that is passed to both IPSs during experiment ranges from 70% to 90% of total network throughput capacity. Number of intrusions in each experiment iteration (48 hours time) is maintained between 100-1000.

Change in throughput does not affect the IPS's capability of detecting and preventing intrusion attempts but the affect of increased throughput is visible in response time of IPS on network events. Increasing value of throughput means longer the time IPS takes to respond the event. This response time was significantly improved when a data processing unit was integrated with each data collection components hardware-based distributed IPS.

When number of intrusions increased from 100 to 1000 with increased throughput, software based IPS fails to identify and prevent almost 4% of network intrusions causing network security compromised. This ratio in hardware-based IPS remains less than 1%. In case of software based IPS number of intrusions in data show inverse relationship with IPS's capability of preventing attacks but for hardware-based IPS this relationship shows a linear behavior.

1. Dataset Reduction: Data being passed to IPSs is well thought and especially planned for this experiment. Data with security threats and intrusion attempts is mixed with clean data to check the detection power of IPSs. Each IPS receives at least 70% of total network throughput capacity with 100 intrusion attempts for average case and 1000 intrusions for extreme case. Data passed includes well known protocols data including TCP, ICMP and UDP etc.
2. Hypothesis Testing: Data produced during the experiment is tested to reject NULL hypothesis. This experiment proved the comparative effectiveness of hardware-based distributed IPS with small processing units integrated into data collecting components over software based IPS. Man-Whitney test is used to compare data produced by both IPS types.
3. Presentation and Package: Experiment is executed successfully and on the basis of findings and extracted facts following can be concluded:
   - Hardware-based distributed IPS (A) is better than software-based centralized IPS (B) both in terms of time (A responded as an average of 20% less time than what B did) and successfully identifying/preventing intrusion attempts (A detected/prevented 99.2% real threats whereas B did only 96.1%).
   - Number of false positives was almost equal (2% in each case).
   - An offers further efficient behavior in terms of time when some processing power is integrated with data collection components.
   - Maintainability is the only thing in which B is better than A (B takes only half the time of what A takes for maintainability and upgrading).

## 4 FOCUS GROUP AS QUALITATIVE APPROACH

A focus group is a tool in which a small group of people make a discussion on chosen topic in an informal setting [1, 15, and 16]. Focus Groups are a good choice for qualitative choice in the field of IPS since it is an emerging field in computer science.

Focus group is a right choice for this study because network security and intrusions are the problems for virtually all organizations. Organizations have to make their resources available online for better communication and con-venience. Security of this information is so critical that its theft, destruction, modification or unauthorized access can cause organizations huge financial [6] and reputation loss [7]. Different organizations showed great interest in this study and security professionals from banks, academic institutions, network managers and security products development teams shared their ideas in this focus group.

Focus group serves in our study as qualitative method. Focus of our study remained on "Enhancement of network security level with help if IPS without causing too much changes in existing infrastructure of networks". Security is so vast topic that it is quite easy to get lost in irrelevant details, so special filtration and scrutiny of data was done to remain focused on our original issue. Study was done without having emphasis on any specific IPS product so that results produced are generic and effective.

### 4.1 Objectives
"*Focus groups are conducted to obtain specific type of information from a clearly identified set of individuals*" [2].

### 4.2 Research Strategy
Research strategy for this study is based on focus groups. This method is best of an emerging field like network security and network intrusions. Also, this study aims to benefit everyone accessing information on computer networks.

### 4.3 Data Collection
According to [8], data collection in qualitative research has six methods. We used for methods (questionnaires, observations, document reviews and interviews) for data collection.

### 4.4 Data Recording
1. Questionnaires: According to David Leigh [9], research questions are of six types (open, closed, leading, reflective, loaded and focused). For this study, mostly open questions are used for possibility of free response and convenience. Some closed questions were also included in questionnaires, where clear agreement or disagreement of respondent was required.

We made many questionnaires available on our online discussion forum (specially developed for this study). Our focus and many independent observers responded to these questionnaires and these responses were stored in databases for analysis.

Following are two of the many questionnaires designed for this study. First questionnaire concerns every network security related professional and second questionnaire is related to only network experts using hardware-based distributed IPS.

(Questionnaire-1 for security personals using any kind of IPS)
1. What kind of organization are you representing?
2. How critical your security needs are?
3. What kind of security attacks do you experience often?
4. How often do you experience outage of service due to intrusion attempts?
5. How have you implemented your IPS on your network?
6. How effective is your IPS against intrusion attempts?
7. What kind of alerts does your IPS generate on intrusion attempt?
8. What are consequences of a successful intrusion attempt on your network?
9. Does your IPS capable of standing against attack on itself?
10. What steps does your IPS take on a successful intrusion attempt?

(Questionnaire-2 for security personals using hardware-based distributed IPS)
1. How many systems are you running on your network?
2. Are you using your network gateway as central point for data collection on your network or multiple points for data collection?
3. How have you implemented your data analysis unit?
4. Have you ever used some other kind of implementation for IPS? If yes, which implementation of IPS (software-based centralized or hardware-based distributed) you find better in your network scenario?
5. How fast is your IPS in detecting hybrid attacks?
6. How effective is your IPS in preventing post intrusion events?
7. What is the number of Known Attack Signatures (KAS) in your IPS database?
8. How often do you update KASs?
9. What is maximum throughput of your network?
10. What is maximum throughput up to which your IPS works perfectly without causing efficiency in real time?

2. Observations: Different groups were given responsibilities to observe certain organizations. One group observed online transaction network for a bank, second group observed one online shopping mall and third group observed a university network. All groups observed the traffic patterns on these networks, different intrusion attempts, sequence of steps taken by respective IPSs to prevent intrusion attempts and logs generated by IPSs. Attack types and post attacks events were very interesting events to observe.

3. Document Reviews: As described earlier, an online discussion forum was developed for this study where responses for all focus groups were published. Important documents that were published on this forum include focus groups discussions summaries, security expert's interviews, related articles and data, questionnaires, observations on networks and statistics of intrusion attempts on networks. All these documents were used in analysis phases later. Some documents that were used in analysis but were not made available online were "network logs".

4. Interviews: "*Interviews are most often used to gather detailed, qualitative descriptions of how programs operate and how stakeholders perceive them*" [8]. During this study, many network security experts were interviewed on different security topics and their views were published on online discussion forum. These interviews were of great importance to know finer de-tails of security technologies.

## 4.5 Data Analysis and Interpretation

After collecting all the data at our discussion forum, data analysis is done by our focus groups. Data is divided into categories and subcategories. Each category is assigned to a focus group and each focus group is responsible for summarization of that category findings.

When each focus group has finished its job data is collected at a central location. Representation from all focus groups conducted detailed discussions and filtered irrelevant details. The agreed upon results are published once again on online discussion forum and once again experts of the field are asked to comment on these results. This way, finer suggestions are accommodated and results are published finally.

## 5 VALIDITY OF RESULTS

Reliability and validity of results in qualitative research cannot be defined in the same way as it is defined in case of quantitative research [11]. In qualitative research, validity and reliability refers to trustworthiness of research results. Results generated by this study are validated through interpretation of observations and comments posted on our online discussion form. These observations are posted in form of interviews contents, suggestions, answers to questionnaires and related articles.

A threat to this validity of this study can be failure of understanding the level of effectiveness of some specific security attack. Professionals may give a little more or a little less importance to a security threat than what it actually deserves.

Another threat to validity can be misunderstanding the context in which that security attack is valid. A security threat out of its context may not be a threat at all or vise versa.

Last threat to this study can be too much specific nature of study. This study is about effectiveness of "Hardware-based distributed IPS". A security expert's comments that is not master to IPS technology, may lead us to not realistic results.

## 6 CONCLUSION

This paper presents a design which includes qualitative and quantitative methods to find-out the improvement area in IPSs. In this paper, two methods are used, Experiment and Focus group. First method represents quantitative study and second represents qualitative study. Usage of both qualitative and quantitative methods in parallel helped us to generalize this study results. Results produced by both methodologies support each other and in this way this study is validated. Further we have explored when any network is under intrusion attack than how we can reduce the responding time for Intrusion prevention system. Also how to handl this attack without effecting speed of network negitavily.

**Muhammad Imran Shafi** completed Master of Science in Computer Science from Blekinge Institute of Technology Sweden in 2008, MSc in Computer Science and BSc from Punjab University Pakistan. Currently he is SEO of SamSoft (software house) in Lahore and visiting Lecturer in Punjab University Pakistan.

**Muhammad Akram** obtained Master of Science in Computer Science from Blekinge Institute of Technology Sweden in 2008, Masters in Computer Science, and BSc from University of Azad Jammu & Kashmir, Pakistan. Currently he is working as a Lecturer in College of Computer Science and Information System, Najran University Saudi Arabia.

**Sikandar Hayat** has obtained Masters in Electrical Engineering with emphasis in telecommunication from Blekinge Institute of Technolo-

gy Sweden in 2009, Master in Information Technology and BSc from University of Azad Jammu & Kashmir Pakistan. Currently he is working as Student Administrator & Project Assistant in IT department of Blekinge Institute of Technology, Sweden.

**Imran Sohail** completed his Master of Science degree from Sweden in 2009 and B.E in 2000. Currently he is working as Assistant professor in Fatima Jinnah University. Before joining FJWU, he was working as a web developer in Eriscsson Stockholm, Sweden. He has vast experience of programming and has developed number of applications.